\documentclass[10pt,a4paper]{amsart}
\usepackage[margin=20mm]{geometry}
\usepackage[numbers]{natbib}
\usepackage[colorlinks]{hyperref}

\usepackage{amssymb,amsmath}
\usepackage{amsthm}
\usepackage[usenames,dvipsnames]{xcolor}
\usepackage{color}

\input xy
\xyoption{all} %\tolerance=500

\numberwithin{equation}{section}

\newtheorem{theorem}{Theorem}[section]

\theoremstyle{definition}
\newtheorem{definition}[theorem]{Definition}
\newtheorem{remark}[theorem]{Remark}
\newtheorem{example}[theorem]{Example}

\newcommand{\rmd}{\textnormal{d}}

\newcommand{\N}{\mathbb{N}}

\DeclareMathOperator{\End}{End}

\DeclareMathOperator{\Der}{Der}

\font\black=cmbx10 \font\sblack=cmbx7 \font\ssblack=cmbx5 \font\blackital=cmmib10  \skewchar\blackital='177
\font\sblackital=cmmib7 \skewchar\sblackital='177 \font\ssblackital=cmmib5 \skewchar\ssblackital='177
\font\sanss=cmss10 \font\ssanss=cmss8 %scaled 900
\font\sssanss=cmss8 scaled 600 \font\blackboard=msbm10 \font\sblackboard=msbm7 \font\ssblackboard=msbm5
\font\caligr=eusm10 \font\scaligr=eusm7 \font\sscaligr=eusm5  \font\fraktur=eufm10
\font\sfraktur=eufm7 \font\ssfraktur=eufm5 
\font\bsymb=cmsy10 scaled\magstep2
\def\all#1{\setbox0=\hbox{\lower1.5pt\hbox{\bsymb
       \char"38}}\setbox1=\hbox{$_{#1}$} \box0\lower2pt\box1\;}
\def\exi#1{\setbox0=\hbox{\lower1.5pt\hbox{\bsymb \char"39}}
       \setbox1=\hbox{$_{#1}$} \box0\lower2pt\box1\;}

\def\tx#1{{\fam0\relax#1}}

\newfam\bifam
\textfont\bifam=\blackital \scriptfont\bifam=\sblackital \scriptscriptfont\bifam=\ssblackital

\newfam\blfam
\textfont\blfam=\black \scriptfont\blfam=\sblack \scriptscriptfont\blfam=\ssblack

\newfam\bbfam
\textfont\bbfam=\blackboard \scriptfont\bbfam=\sblackboard \scriptscriptfont\bbfam=\ssblackboard

\newfam\ssfam
\textfont\ssfam=\sanss \scriptfont\ssfam=\ssanss \scriptscriptfont\ssfam=\sssanss
\def\sss#1{{\fam\ssfam\relax#1}}

\newfam\clfam
\textfont\clfam=\caligr \scriptfont\clfam=\scaligr \scriptscriptfont\clfam=\sscaligr

\newfam\frfam
\textfont\frfam=\fraktur \scriptfont\frfam=\sfraktur \scriptscriptfont\frfam=\ssfraktur

\def\hpb#1{\setbox0=\hbox{${#1}$}
    \copy0 \kern-\wd0 \kern.2pt \box0}
\def\vpb#1{\setbox0=\hbox{${#1}$}
    \copy0 \kern-\wd0 \raise.08pt \box0}

\def\pmb#1{\setbox0\hbox{${#1}$} \copy0 \kern-\wd0 \kern.2pt \box0}
\def\pmbb#1{\setbox0\hbox{${#1}$} \copy0 \kern-\wd0
      \kern.2pt \copy0 \kern-\wd0 \kern.2pt \box0}
\def\pmbbb#1{\setbox0\hbox{${#1}$} \copy0 \kern-\wd0
      \kern.2pt \copy0 \kern-\wd0 \kern.2pt
    \copy0 \kern-\wd0 \kern.2pt \box0}
\def\pmxb#1{\setbox0\hbox{${#1}$} \copy0 \kern-\wd0
      \kern.2pt \copy0 \kern-\wd0 \kern.2pt
      \copy0 \kern-\wd0 \kern.2pt \copy0 \kern-\wd0 \kern.2pt \box0}
\def\pmxbb#1{\setbox0\hbox{${#1}$} \copy0 \kern-\wd0 \kern.2pt
      \copy0 \kern-\wd0 \kern.2pt
      \copy0 \kern-\wd0 \kern.2pt \copy0 \kern-\wd0 \kern.2pt
      \copy0 \kern-\wd0 \kern.2pt \box0}

\mathchardef\za="710B  %\alpha
\mathchardef\zb="710C  %\beta
\mathchardef\zg="710D  %\gamma
\mathchardef\zd="710E  %\delta
\mathchardef\zve="710F %\epsilon
\mathchardef\zz="7110  %\zeta
\mathchardef\zh="7111  %\eta
\mathchardef\zvy="7112 %\theta
\mathchardef\zi="7113  %\iota
\mathchardef\zk="7114  %\kappa
\mathchardef\zl="7115  %\lambda
\mathchardef\zm="7116  %\mu
\mathchardef\zn="7117  %\nu
\mathchardef\zx="7118  %\xi
\mathchardef\zp="7119  %\pi
\mathchardef\zr="711A  %\rho
\mathchardef\zs="711B  %\sigma
\mathchardef\zt="711C  %\tau
\mathchardef\zu="711D  %\upsilon
\mathchardef\zvf="711E %\phi
\mathchardef\zq="711F  %\chi
\mathchardef\zc="7120  %\psi
\mathchardef\zw="7121  %\omega
\mathchardef\ze="7122  %\varepsilon
\mathchardef\zy="7123  %\vartheta
\mathchardef\zf="7124  %\varomega
\mathchardef\zvr="7125 %\varrho
\mathchardef\zvs="7126 %\varsigma
\mathchardef\zf="7127  %\varphi
\mathchardef\zG="7000  %\Gamma
\mathchardef\zD="7001  %\Delta
\mathchardef\zY="7002  %\Theta
\mathchardef\zL="7003  %\Lambda
\mathchardef\zX="7004  %\Xi
\mathchardef\zP="7005  %\Pi
\mathchardef\zS="7006  %\Sigma
\mathchardef\zU="7007  %\Upsilon
\mathchardef\zF="7008  %\Phi
\mathchardef\zW="700A  %\Omega
\mathchardef\zC="7009  %\Psi

\newcommand{\be}{\begin{equation}}
\newcommand{\ee}{\end{equation}}

\newcommand{\bea}{\begin{eqnarray}}
\newcommand{\eea}{\end{eqnarray}}
\def\*{{\textstyle *}}
\newcommand{\R}{{\mathbb R}}

\newcommand{\Z}{{\mathbb Z}}

\newcommand{\s}{{\textstyle *}}

\newcommand{\A}{{\mathcal A}}

\def\Sec{\sss{Sec}}

\def\sT{{\sss T}}

\def\xi{\tx{i}}

\def\s*{{\scriptstyle *}}

\def\cO{\mathcal{O}}

\newcommand{\beas}{\begin{eqnarray*}}
\newcommand{\eeas}{\end{eqnarray*}}
\title{The graded differential geometry of mixed symmetry tensors}
\author{Andrew James Bruce \& Eduardo Ibarguengoytia}
\address{Mathematics Research Unit, University of Luxembourg, Maison du 	 Nombre 6, avenue de la Fonte,  L-4364 Esch-sur-Alzette, Luxemburg}
\email{andrewjamesbruce@googlemail.com, ~ eduardo.ibarguengoytia@uni.lu}

\date{\today}
\begin{document}

\begin{abstract}
We show how the theory of $\Z_2^n$-manifolds - which are a non-trivial generalisation of supermanifolds - may be useful in a  geometrical approach to mixed symmetry tensors such as the dual graviton. The geometric aspects of such tensor fields on both flat and curved space-times are discussed.\par
\smallskip\noindent
\begin{tabular}{ll}
{\bf Keywords:} & $\Z_2^n$-manifolds;~mixed symmetry tensors;~dual gravitons\\
{\bf MSC 2010:}& 53C80;~ 58A50;~83C65 \\
{\bf PACS Numbers:} &  02.40.Gh;~ 04.20.Cv;~ 11.10.kk 
\end{tabular}
\end{abstract}

 \maketitle

\vspace{-20pt}
\setcounter{tocdepth}{2}
 % \tableofcontents

\section{Introduction}
Recall that differential forms are covariant tensor fields that are completely antisymmetric in their indices. Furthermore, it is  well-known that supermanifolds offer a convenient set-up in which to deal with differential forms.  In particular, differential forms can be understood as functions on the supermanifold $\Pi \sT M$ known as the antitangent bundle. This supermanifold is constructed by taking the tangent bundle of a manifold and then declaring the fibre coordinates to be Grassmann odd. Moreover, the antitangent bundle canonically comes equipped with an odd vector field which `squares to zero', this vector field is identified with the de Rham differential (see for example Vaintrob \cite{Vaintrob:1996} for details).  Symmetric forms are covariant tensor fields that are completely symmetric in their indices and can be understood as polynomial functions on the tangent bundle of the manifold under study. There is no  symmetric analogue of the de Rham differential on an arbitrary smooth manifold unless one invokes an affine connection. Mixed symmetry tensor fields are covariant tensors fields with more than one set of antisymmetrised indices. Mixed symmetry tensor fields represent a  natural generalisation of differential forms in which the tensors are  neither fully symmetric nor antisymmetric. From the perspective of differential geometry, mixed symmetry tensors are not well studied.  From a representation theory point of view, they correspond to Young diagrams with more than one column. In physics, such tensor fields appear in the context of higher spin fields, dual gravitons, double dual gravitons etc. as found in various formulations of supergravity  and string theory. In particular, the particle spectrum of string theory contains beyond the massless particles of the effective supergravity theory, an infinite tower of massive particles of  higher and higher  spin.  Thus, if one wants to consider the theory beyond the effective supergravity theory, one is forced to contend with mixed symmetry tensors. Furthermore, Hull \cite{Hull:2000, Hull:2001} suggested that dual gravitons and double dual gravitons play a fundamental r\^ole  in the electromagnetic duality of gravitational theories. Alongside this, mixed symmetry tensors naturally appear in dual double theory  \cite{Bergshoeff:2016} and  it is known that in string theory certain mixed symmetry tensors couple to exotic branes \cite{Chatzistavrakidis:2014}.  To our knowledge, the first study of mixed symmetry tensor fields was Curtright \cite{Curtright:1985} who studied a generalised version of gauge theory. For a review of mixed symmetry tensors, including some historical remarks, the reader may consult Campoleoni \cite{Campoleoni:2010}. Recently,  Chatzistavrakidis \emph{et al}. \cite{Chatzistavrakidis:2017} showed how to reformulate  Galileon action functionals in an index-free framework using a generalised notion of a supermanifold. The reader should also note that these results are part of Khoo's PhD dissertation \cite{Khoo:2016}. Their theory involves two sets of Grassmann variables that mutually commute. However, assigning a degree of one to \emph{all} the Grassmann variables does not lead to a consistent notion of a ``generalised supermanifold''. For one,  the commutation rules of the coordinates are not defined  by their degree. Thus,  it is impossible to make global sense of the geometry: what is the commutation rule for two arbitrary degree one functions?  These difficulties are cured by using a bi-grading and the theory of  $\Z_2^n$-manifolds with $n=2$.  Moreover, the  formalism of bi-forms (and multi-forms) as developed by Dubois-Violette \& Henneaux \cite{Dubois-Violette:2002}, de Medeiros \& Hull \cite{deMedeiros:2003}, and  Bekaert \& Boulanger   \cite{Bekaert:2004}, is naturally accommodated within this framework. \par
The locally ringed space approach to $\Z_2^n$-manifolds is currently work in progress initially started by  Covolo \emph{et al}. \cite{Covolo:2016,Covolo:2016a,Covolo:2016b}. However, with the basic tenets in place, the time is ripe to seek applications and links with known constructions.  Very loosely, $\Z_2^n$-manifolds are `manifolds' in which we have $\Z_2^n$-graded, $\Z_2^n$-commutative coordinates. The sign rules are controlled by the standard scalar product on $\Z_2^n$. Hence, in general, we have sets of coordinates that anti-commute  amongst themselves while commuting across the sets. This is exactly what we require in order to describe mixed symmetry tensors. The one complication is that, in general, there are also formal coordinates that are not nilpotent. This means that we must consider formal power series and not just polynomials in the formal coordinates. However, with the applications to mixed symmetry tensors in mind, we will not need to dwell on this subtlety. We will concentrate on mixed tensors with two `blocks' of antisymmetric indices and so we will  employ very particular $\Z_2^2$-manifolds, for the most part with no non-nilpotent formal coordinates. \par 
We liken the current situation to the early days of supersymmetry and in particular the initial works on superspace methods. In particular, physicists worked rather formally with commuting and anticommuting coordinates largely unaware of that the mathematical theory of supermanifolds was concurrently being developed in the Soviet Union by Berezin and collaborators. We speculate that  $\Z_2^n$-manifolds will shed light  on various aspects of theoretical physics and here we suggest just one potentially useful facet.\par 
\medskip
\noindent \textbf{Arrangement:} In section \ref{sec:Basics} we present the very basics of the theory of $\Z_2^n$-manifolds needed for the rest of this paper. We then proceed in section \ref{sec:Minkowski} to discuss how to use $\Z_2^2$-manifolds to understand bi-forms over Minkowski space-time. It is shown that the algebra of bi-forms over Minkowski space-time comes canonically equipped with a pair of de Rham differentials.  Generalising the constructions to the setting of curved space-times is the subject of section \ref{sec:Curved}. In particular, the analogues of the de Rham differentials require the use of the Levi-Civita connection due to the non-fully antisymmetric nature of bi-forms. This means that in general, we have a pair of `non-homological vector fields' and cannot construct a genuine bi-complex. However, such vector fields still define infinitesimal diffeomorphisms that we interpret as `supersymmetries'. In section \ref{sec:VBvalued} we show how to extend our formalism to include bi-forms that take their values in a vector bundle. For instance, this leads to the notion of twisted bi-forms where the vector bundle is the density bundle on the curved space-time. We conclude in section \ref{sec:Concl} with some remarks.

%%%%%%%%%%%%%%%%%%%%%%%%%%%%%%%%%%%%%%%%%%%%%%%%%%%%%%

\section{Basics of $\Z_2^n$-geometry}\label{sec:Basics}
The first reference to $\Z_{2}^{n}$-manifolds (\emph{coloured manifolds}) is Molotkov  \cite{Molotkov:2010}  who developed a functor of points approach. The locally ringed space approach to $\Z_{2}^{n}$-manifolds is presented in   \cite{Covolo:2016}.  We will draw upon this  heavily and not present proofs of any formal statements. We work over the field $\R$ and in our notation $\Z_2^n := \Z_2 \times \Z_2 \times \Z_2$ ($n$-times). A \emph{$\Z_2^n$-graded algebra} is an $\R$-algebra  with a decomposition into vector spaces $\A :=  \oplus_{\gamma \in \Z_2^n} \A_\gamma$, such that the multiplication respect the $\Z_2^n$-grading, i.e., $\A_\alpha \cdot \A_\beta \subset \A_{\alpha + \beta}$. Furthermore, we will always assume the algebras to be associative and  unital. If for any pair of homogeneous elements $a \in \A_\alpha$ and $b \in \A_\beta$ we have that 
\begin{equation}\label{eq:Z2ncommrules}
a \cdot b = (-1)^{\langle \alpha , \beta\rangle } b \cdot a,
\end{equation}
where $\langle -, -\rangle$ is the standard scalar product on $\Z_2^n$, then we have a  \emph{$\Z_2^n$-commutative algebra}.\par 
The basic objects we will employ are smooth $\Z_2^n$-manifolds. Essentially, such objects are `manifolds' equipped with both standard commuting coordinates and formal coordinates of non-zero $\Z_2^n$-degree that $\Z_2^n$-commute according to the general sign rule \eqref{eq:Z2ncommrules}. Note that in general - and in stark contrast to the $n=1$ case of supermanifolds -  we have formal coordinates that are \emph{not} nilpotent. \par 
In order to keep track of the various formal coordinates, we need to introduce a convention on how we fix the order of elements in $\Z_{2}^{n}$, we do this \emph{lexicographically}. For example, with this choice of ordering
$$\Z_{2}^{2} = \{ (0,0),  \: (0,1), \: (1,0), \: (1,1)\}\,.$$
Note that other choices of ordering have appeared in the literature. A tuple $\mathbf{q} = (q_{1}, q_{2}, \cdots , q_{N})$, where  $N = 2^{n}-1$ provides \emph{all} the information about the formal coordinates. We can now recall the definition of a $\Z_2^n$-manifold.
\begin{definition}
A (smooth) $\Z_{2}^{n}$-\emph{manifold} of dimension $p |\mathbf{q}$ is a locally $\Z_{2}^{n}$-ringed space $ \mathcal{M} := \left(M, \cO_M \right)$, which is locally isomorphic to the $\Z_{2}^{n}$-ringed space $\mathbb{R}^{p |\mathbf{q}} := \left( \mathbb{R}^{p}, C^{\infty}_{\mathbb{R}^{p}}[[\zx]] \right)$. Local sections of $M$ are formal power series in the $\Z_{2}^{n}$-graded variables $\zx$ with  smooth coefficients,
$$\cO_M(U) \simeq C^{\infty}(U)[[\zx]] :=  \left \{ \sum_{\hat\alpha \in \mathbb{N}^{N}}^{\infty}  \zx^{\hat \alpha}f_{\hat\alpha} ~ | \: f_{\hat \alpha} \in C^{\infty}(U)\right \},$$
for `small enough' open domains $U\subset M$.   \emph{Morphisms} between $\Z_{2}^{n}$-manifolds are  morphisms of $\Z_{2}^{n}$-ringed spaces, that is,  pairs $\Phi = (\phi, \phi^{*}) : (M, \cO_M) \rightarrow  (N, \cO_N)$ consisting of  a continuous map  $\phi: M \rightarrow N$ and sheaf morphism $\phi^{*} : \cO_N \rightarrow \cO_M$, i.e., a family of $\Z_2^n$-algebra morphisms $\phi^*_{V}: \cO_N(V) \rightarrow \cO_M(\phi^{-1}(V))$, where $V \subset N$  is open. We will refer to the global sections of the structure sheaf $\cO_M$ as \emph{functions} on $M$ and denote them as $C^{\infty}(\mathcal{M}) := \cO_{M}(M)$.
\end{definition}
\begin{example}[The local model]\label{exp:SuperDom}
The locally $\Z_{2}^{n}$-ringed space $\mathcal{U}^{p|\mathbf{q}} :=  \big(\mathcal{U}^p , C^\infty_{\mathcal{U}^p}[[\zx]] \big)$, where $\mathcal{U}^p \subseteq \R^p$ is naturally a $\Z_2^n$-manifold -- we refer to such $\Z_2^n$-manifolds as \emph{$\Z_2^n$-superdomains} of dimension $p|\mathbf{q}$.  We can employ (natural) coordinates $(x^a, \zx^\alpha)$ on any $\Z_2^n$-superdomain, where $x^a$ form a coordinate system on $\mathcal{U}^p$ and the $\zx^\alpha$ are formal coordinates. 
\end{example} 
Many of the standard results from the theory of supermanifolds pass over to $\Z_{2}^{n}$-manifolds. For example, the topological space $M$ comes with the structure of a smooth manifold of dimension $p$, hence our suggestive notation. Moreover, there exists a canonical projection $\epsilon : \cO(M) \rightarrow C^{\infty}(M)$. What makes $\Z_{2}^{n}$-manifolds a very workable form of noncommutative geometry is the fact that we have well-defined local models. Much like the theory of manifolds, one can construct global geometric concepts via the glueing of local geometric concepts. That is, we can consider a $\Z_{2}^{n}$-manifold as being cover by  $\Z_2^n$-superdomains together with specified glueing information given by coordinate transformations, composed by homomorphisms $$\Psi_{\beta\alpha}:=\Psi_\beta ^{-1}\Psi_\alpha:\Psi_{\alpha}^{-1}(\Psi_\alpha(U_\alpha)\cap\Psi_\beta(U_\beta))\rightarrow \Psi_{\beta}^{-1}(\Psi_\alpha(U_\alpha)\cap\Psi_\beta(U_\beta)) ,$$
which are labelled by the different local models $(U_\alpha,C^{\infty}(U_\alpha )[[\zx]])$, $\{\Psi_\alpha:U_\alpha \rightarrow\ \Psi_\alpha(U_\alpha)\subset M\}$, whenever $U_\alpha \cap U_\beta \ne \emptyset$; and a graded unital $\mathbb{R}-$algebra morphism $\Psi_{\beta\alpha}^*: C^{\infty}(U_\beta )[[\zx^\prime]]\longrightarrow C^{\infty}(U_\alpha )[[\zx]]$.

We have the \emph{chart theorem} (\cite[Theorem 7.10]{Covolo:2016}) that basically says that morphisms between $\Z_{2}^{n}$-superdomains can be completely described by local coordinates and that these local morphisms  can then be extended uniquely to morphisms of locally $\Z_{2}^{n}$-ringed spaces. This allows one to proceed to describe the theory much as one would on a standard smooth manifold in terms of local coordinates. Indeed, we will employ the standard abuses of notation when dealing with coordinate transformations and morphisms. In particular,  the explicit way of computing change of coordinates concerning any geometrical object are well understood and work identically as in classical differential geometry. In essence, one need only take into account that $\Z_2^n$-degree needs to be preserved under any permissible changes of coordinates. For example, \emph{vector fields} are defined as $\Z_2^n$-graded derivations of the global sections, $X\in \Der(C^{\infty}(\mathcal{M}) \subset \End(C^\infty(\mathcal{M}))$, that are compatible with restrictions. That is, given some open subset $U \subset M$, we can always `localise' the vector field, i.e.,   $X\vert _U=X_U\in \Der(\cO_M(U))$. Furthermore, if this open is `small enough', we can employ local coordinates $(x^a, \zx^\alpha)$ and write 
$$X_U = X ^{a}(x, \zx) \frac{\partial}{\partial{x ^{a}}} + X^\alpha(x,\zx) \frac{\partial}{\partial \zx^\alpha}\,.$$ 
Under changes of local coordinates 
\begin{align*}
& x^{a'} = x^{a'}(x, \zx), && \zx^{\alpha'} = \zx^{\alpha'}(x, \zx), 
\end{align*}
remembering the abuses of notation and that $\Z_2^n$-degree is preserved, the induced transformation law on the components of the vector field follow from the chain rule and are given by
\begin{align*}
& X^{a'} = X^b \frac{\partial x^{a'}}{\partial x^b} + X^\beta \frac{\partial x^{a'}}{\partial \zx^\beta}, && X^{\alpha'} = X^b \frac{\partial \zx^{\alpha'}}{\partial x^b} + X^\beta \frac{\partial \zx^{\alpha'}}{\partial \zx^\beta}. 
\end{align*}
See Covolo \emph{et al.} \cite[Lemma 2.2]{Covolo:2016b} for details. The reader can easily verify that the $\Z_2^n$-graded commutator of two vector fields is again a vector field and that the obvious $\Z_2^n$-graded version of the Jacobi identity holds. \par
As is customary in classical differential geometry, we will not write out the restrictions of geometric objects explicitly and simply write objects in terms of there components in some chosen local coordinate system.  In other words, one can work locally on $\Z_2^n$-manifolds in more-or-less the same way as one works on classical manifolds and indeed, supermanifolds.  The glaring exception here is the theory of integration on  $\Z_2^n$-manifolds which is expected to be quite involved (see Poncin \cite{Poncin:2016} for work in this direction).

%%%%%%%%%%%%%%%%%%%%%%%%%%%%%%%%%%%%%%%%%%%%%%%%%%%%%%

\section{Mixed symmetry tensors over Minkowski space-time}\label{sec:Minkowski}
Consider $D$-dimensional Minkowski space-time $M = (\R^D, \eta)$. The Poincar\'{e} transformations we write as 
$$x^\mu \mapsto x^{\mu'} =  x^\nu \Lambda_{\nu}^{\:\: \mu'}  ~+~ a^{\mu'}\,. $$
We now wish to construct a $\Z_2^2$-manifold built from $M$ in a canonical way. In particular, consider
$$\mathcal{M} := \sT M[(0,1)] \times_M \sT M[(1,0)],$$
where we have indicated the assignment of the $\Z_2^2$-grading to the fibre coordinates on each tangent bundle. It is straightforward to see that we do indeed obtain a $\Z_2^2$-manifold in this way by using coordinates (see \cite[Proposition 6.1]{Covolo:2016}). Specifically, we can always employ (global) coordinates of the form
$$\big( \underbrace{x^\mu}_{(0,0)}, ~ \underbrace{\zx^\nu}_{(0,1)}, ~ \underbrace{\theta^\rho}_{(1,0)}   \big)\,,$$ 
where we have signalled the assignment of $\Z_2^2$-grading. Note that we have the  non-trivial $\Z_2^2$-commutation rules
\begin{align*}
& \zx^\mu \zx^\nu = {-} \zx^\nu \zx^\mu, && \theta^\mu \theta^\nu = {-} \theta^\nu \theta^\mu, && \zx^\mu \theta^\nu = {+} \theta^\nu \zx^\mu.
\end{align*}
Thus, while each `species' of non-zero degree coordinate are themselves nilpotent,  across `species' they commute. This is, of course, very different from the case of standard supermanifolds.  The Poincar\'{e} transformations induce the obvious linear coordinate transformations on the formal coordinates
\begin{align*}
\zx^{\nu'} = \zx^{\nu}\Lambda_{\nu}^{\:\: \nu'}, && \theta^{\rho'} = \theta^{\rho}\Lambda_{\rho}^{\:\: \rho'}.
\end{align*}
Clearly, these transformation laws respect the assignment of $\Z_2^2$-grading and satisfy (rather trivially) the cocycle condition. Thus, we do indeed obtain a $\Z_2^2$-manifold in this way.  As the coordinate transformations respect the obvious bundle structure and do not `mix'  the non-zero degree coordinates we have an example of a so-called \emph{split $\Z_2^2$-manifold} \cite{Covolo:2016a}. The fact that we do not, in this case, have non-zero degree coordinates that are not nilpotent means that we only deal with polynomials in the formal coordinates. \par 
The space of $(p,q)$-forms on $M$ we define as 
$$\Omega^{(p,q)}(M) :=  C^\infty(\mathcal{M})_{(p,q)},$$
where we naturally have the $\N\times \N$-grading given by the polynomial order in each formal coordinate. As we are considering linear coordinate changes only, this order is well-defined.  By considering all  possible degrees we obtain a unital $\Z_2^2$-commutative algebra
$$\Omega(M) :=   C^\infty(\mathcal{M}) = \bigoplus_{(p,q) \in \N\times \N}^{(D,D)} \Omega^{(p,q)}(M),$$
which we refer to as the algebra of \emph{bi-forms}: which we can view as the algebra of `differential forms with values in differential forms'. Note that we naturally have a $C^\infty(M) = \Omega^{(0,0)}(M)$ module structure on the space of all bi-forms. \par 
In coordinates, any $(p,q)$-form can be written as
$$\omega^{(p,q)}(x, \zx, \theta) = \frac{1}{p! q!} \: \theta^{\nu_1} \cdots \theta^{\nu_p} \zx^{\mu_1} \cdots \zx^{\mu_q}  \: \omega_{\mu_q \cdots \mu_1 | \nu_q \cdots \nu_1}(x).$$
Due to the $\Z_2^2$-commutation rules, we have the relation that $\omega_{[\mu_q \cdots \mu_1] |[\nu_q \cdots \nu_1]} = \omega_{\mu_q \cdots \mu_1 |\nu_q \cdots \nu_1}$ and $\omega_{[\mu_q \cdots \mu_1]| [\nu_q \cdots \nu_1]} = \omega_{[\nu_q \cdots \nu_1]| [\mu_q \cdots \mu_1]}$  Note that we will not insist on any further relations in general. \par 
\begin{example}
The dual graviton in $D$-dimensions is a $(1,D-3)$-form and so is given in coordinates as
$$\textnormal{C}(x, \zx, \theta) = \frac{1}{(D-3)!}  \: \theta^\nu \zx^{\mu_1} \cdots \zx^{\mu_{D-3}} \: C_{\mu_{D-3} \cdots \mu_1 | \nu}(x).$$
Similarly, the double dual graviton in $D$-dimensions of a $(D-3, D-3)$-form and so is given in coordinates as
$$ \textnormal{D}(x, \zx, \theta) = \frac{1}{(D-3)! (D-3)!} \:  \theta^{\nu_1} \cdots \theta^{\nu_{D-3}} \zx^{\mu_1} \cdots \zx^{\mu_{D-3}} \: D_{\mu_{D-3} \cdots \mu_1 | \nu_{D-3} \cdots \nu_1}(x).$$
See Hull \cite{Hull:2000, Hull:2001} for details of the r\^{o}le of dual gravitons and double dual gravitons in  electromagnetic duality of gravitational theories.  
\end{example}

Canonically, the algebra of bi-forms on $D$-dimensional Minkowski space-time comes equipped with a pair of de Rham differentials. These differentials we consider as homological vector fields on the $\Z_2^2$-manifold $\mathcal{M}$. That is, they `square to zero', i.e., $2 \rmd^2 = [\rmd,\rmd] =0$. In coordinate we have
\begin{align*}
\rmd_{(0,1)} = \zx^\mu \frac{\partial}{\partial x^\mu}, && \rmd_{(1,0)} = \theta^\mu \frac{\partial}{\partial x^\mu}\,.
\end{align*}
It is important to note that do indeed have a pair of vector fields in this way.  In particular, the partial derivatives change under Poincar\'{e} transformations as
\begin{align*}
\frac{\partial}{\partial x^{\mu'}} = \Lambda_{\mu'}^{\:\:\mu}\frac{\partial}{\partial x^{\mu}}, && \frac{\partial}{\partial \zx^{\nu'}} = \Lambda_{\nu'}^{\:\:\nu}\frac{\partial}{\partial \zx^{\nu}}, && \frac{\partial}{\partial \theta^{\rho'}} = \Lambda_{\rho'}^{\:\:\rho}\frac{\partial}{\partial \theta^{\rho}}\,.
\end{align*}
Thus, the pair of de Rham differentials are well-defined. It is also clear that they $\Z_2^2$-commute, i.e,
$$[\rmd_{(1,0)}, \rmd_{(0,1)}] := \rmd_{(1,0)} \circ \rmd_{(0,1)}~ {-}~\rmd_{(0,1)} \circ \rmd_{(1,0)} =0\,. $$
In this way, we obtain a \emph{de Rham bi-complex}. Also, note that the interior product and Lie derivative  can also be directly `doubled'. \par 
Canonically we also have  a pair of vector fields of $\Z_2^2$-degree $(1,1)$, given by
\begin{align*}
\Delta_{(0,1)} = \zx^\mu \frac{\partial}{\partial \theta^\mu}, && \Delta_{(1,0)} = \theta^\nu \frac{\partial}{\partial \zx^\nu}\,.
\end{align*}
A direct calculation shows that the non-trivial $\Z_2^2$-commutators are
\begin{align*}
[\Delta_{(0,1)}, \rmd_{(1,0)}] = \rmd_{(0,1)}, && [\Delta_{(1,0)}, \rmd_{(0,1)}] = \rmd_{(1,0)} \,.
\end{align*}
Rather conveniently, we can understand the metric as a $(1,1)$-form and the inverse of the metric as a second-order differential operator given by
\begin{align*}
&\eta := \theta^\mu \zx^\nu \eta_{\nu \mu},
&&\eta^{-1} :=  \eta^{\mu \nu}\frac{\partial^2}{\partial \zx^\nu \partial \theta^\mu},
\end{align*}
respectively. 
\begin{example}
Consider the  Curtright field on $D=5$ Minkowski space-time \cite{Curtright:1985}. Such a field is understood to be the electromagnetic dual of the graviton field. In our language, the Curtright field is an example of a $(1,2)$-form and as such can be written in coordinates as
$$\textnormal{C}(x, \zx, \theta) = \frac{1}{2!} \theta^\rho \zx^\nu \zx^\mu C_{\mu \nu |\rho}(x)\,.$$
There is a further symmetry condition on the Curtright field, i.e., $C_{\mu \nu |\rho} + C_{ \rho \mu | \nu} + C_{\nu \rho| \mu} =0$, which comes from wanting an irreducible representation of the Poincar\'{e} group. This condition can be expressed as
$$\Delta_{(0,1)} \textnormal{C} =  \frac{1}{2! 3} \zx^\rho \zx^\nu \zx^\mu \big(C_{\mu \nu |\rho} + C_{ \rho \mu |\nu} + C_{\nu \rho |\mu} \big) =0.$$
Furthermore, a direct calculation shows that 
 $$ \textnormal{F} := \rmd_{(0,1)}\textnormal{C} = \frac{1}{3!} \theta^\rho \zx^\nu \zx^\mu \zx^\lambda\left( \frac{\partial C_{\mu \nu | \rho}}{\partial x^\lambda} + \frac{\partial C_{ \nu \lambda | \rho}}{\partial x^\mu}  + \frac{\partial C_{  \lambda\nu |\rho}}{\partial x^\nu}\right) 
   =  \frac{1}{3!} \theta^\rho \zx^\nu \zx^\mu \zx^\lambda F_{\lambda \mu \nu |\rho}(x)\,,$$
which we recognise (up to possible  conventions) to be the \emph{Curtright field strength}. Applying $\rmd_{(1,0)}$ to the Curtright field strength yields
$$ \textnormal{E} := \rmd_{(1,0)}\left( \rmd_{(0,1)} \textnormal{C}\right) = \frac{1}{2! 3!} \theta^\omega \theta^\rho \zx^\nu \zx^\mu \zx^\lambda \left( \frac{\partial F_{\lambda \mu \nu | \rho} }{\partial x^\omega}  {-} \frac{\partial F_{\lambda \mu \nu |\omega} }{\partial x^\lambda}  \right) = \frac{1}{2! 3!} \theta^\omega \theta^\rho \zx^\nu \zx^\mu \zx^\lambda E_{\lambda \mu \nu | \rho\omega }(x)\,,$$
which we recognise (up to possible conventions) to be the \emph{Curtright curvature tensor}, which is fully gauge invariant, see Bekaert, Boulanger \&   Henneaux \cite{Bekaert:2003} for details.  Similarly the \emph{Curtright--Ricci tensor} and its trace (again, up to conventions) can be constructed by applying the inverse metric, i.e.,
\begin{align*}
& \eta^{-1}(\textnormal{E}) = \frac{1}{2!}\theta^\rho \zx^\mu \zx^\lambda \eta^{\omega \nu}E_{\lambda \mu \nu | \rho \omega}(x) = \frac{1}{2!}\theta^\rho \zx^\mu \zx^\lambda E_{\lambda \mu |\rho}(x),\\
& \eta^{-1}\big( \eta^{-1}(\textnormal{E})\big) = \zx^\lambda \eta^{\rho \mu} E_{\lambda \mu |\rho}(x)  = \zx^\lambda E_\lambda(x)\,.
\end{align*}
\end{example}

\begin{remark}
The procedure to describe mixed symmetry tensors with more antisymmetric `blocks' is clear.  In particular, if we have $n$ such blocks, then we should consider the $\Z_2^n$-manifold
$$\mathcal{M} := \sT M[(0,\cdots, 0,1)]\times_M\sT M[(0,\cdots, 0,1,0)]\times_M \cdots \times_M \sT M[(1,\cdots, 0,0)]\,,$$
where we have signalled the $\Z_2^n$-degree of the fibre coordinates. Note that we have a canonical de Rham differential in each sector. Thus, the previous statements of this section can be generalised verbatim.
\end{remark}

\begin{remark}
The reader should note that a $\Z_2^n$-grading together with the standard scalar product is enough to encode arbitrary sign rules for finitely generated algebras \cite[Theorem 2.1]{Covolo:2016}. Thus, even more exotic tensors can be encoded using $\Z_2^n$-manifolds. For example, tensors with commuting `blocks' of indices that across `blocks' anticommute can also naturally be formulated in the current setting. We will however, not discuss this further here. 
\end{remark}

%%%%%%%%%%%%%%%%%%%%%%%%%%%%%%%%%%%%%%%%%%%%%%%%%%%%%%

\section{Mixed symmetry tensors over curved space-times}\label{sec:Curved}
Directly extending the constructions to curved space-times $(M, g)$  is not possible. This was for sure noticed in \cite{Chatzistavrakidis:2017}, albeit with no reference to $\Z_2^n$-manifolds. The two de Rham differentials cannot be na\"{\i}vely be considered as vector fields on $\mathcal{M} = \sT M[(0,1)]\times_M \sT M[(1,0)]$. The resolution to this problem is the standard one: we use the Levi-Civita connection to lift the vector fields. The $\Z_2^2$-manifold $\mathcal{M}$ comes equipped with natural coordinates  
$$\big( \underbrace{x^\mu}_{(0,0)}, ~ \underbrace{\zx^\nu}_{(0,1)}, ~ \underbrace{\theta^\rho}_{(1,0)}   \big)\,,$$ 
where again we have signalled the assignment of $\Z_2^2$-grading. The permissible changes of local coordinates are 
\begin{align*}
& x^{\mu'} = x^{\mu'}(x), && \zx^{\nu'} = \zx^\nu \frac{\partial x^{\nu'}}{\partial x^\nu}, && \theta^{\rho'} =  \theta^\rho\frac{\partial x^{\rho'}}{\partial x^\rho}.
\end{align*}
\begin{example}
The covariant Weyl curvature tensor  and covariant Riemannian curvature on any (pseudo-)Riemannian manifold are examples of $(2,2)$-forms.
\end{example}
As standard, we define a covariant derivative 
$$\nabla_\mu :=  \frac{\partial}{\partial x^\mu} {-} \zx^\nu \Gamma_{\nu \mu}^{\rho} \frac{\partial}{\partial \zx^\rho} {-}  \theta^\nu \Gamma_{\nu \mu}^{\rho} \frac{\partial}{\partial \theta^\rho},$$
where $\Gamma_{\nu \mu}^{\rho}$ are the Christoffel symbols of the Levi-Civita connection.  We then define the \emph{covariant de Rham derivatives} as
\begin{align*}
& \nabla_{(0,1)} := \zx^\mu \nabla_\mu = \zx^\mu \frac{\partial}{\partial x^\mu } {-} \zx^\mu\theta^\nu \Gamma_{\nu \mu}^{\rho} \frac{\partial}{\partial \theta^\rho}\,,
&& \nabla_{(1,0)} := \theta^\mu \nabla_\mu = \theta^\mu \frac{\partial}{\partial x^\mu } {-} \zx^\mu\theta^\nu \Gamma_{\nu \mu}^{\rho} \frac{\partial}{\partial \zx^\rho}\,,
\end{align*}
remembering that the Christoffel symbols are symmetric in the lower indices, i.e., the  Levi--Civita connection is torsion free. Due to the transformation rules for the Christoffel symbols both these covariant de Rham derivatives are well-defined vector fields on $\mathcal{M}$. However, in general, we lose the fact that these vector fields are homological and that they commute.  This is in stark contrast to the case of standard differential forms where the covariant derivative (with respect to \emph{any} torsionless connection) reduces to the de Rham differential. Direct calculation shows that
\begin{align*}
[\nabla_{(0,1)}, \nabla_{(0,1)}] & =  R_{(0,1)} = \theta^\mu \zx^\lambda \zx^\nu R_{~\mu \nu \lambda}^\rho(x) \frac{\partial}{\partial \theta^\rho}\,,\\
[\nabla_{(1,0)}, \nabla_{(1,0)}] & =  R_{(1,0)} = \zx^\mu \theta^\lambda \theta^\nu R_{~\mu \nu \lambda}^\rho(x) \frac{\partial}{\partial \zx^\rho}\,,\\
[\nabla_{(1,0)}, \nabla_{(0,1)}] & =  R_{(1,1)} = \zx^{\mu}\theta^{\lambda}\theta^{\nu}R_{~\mu \nu \lambda}^{\rho} \frac{\partial}{\partial \theta^{\rho}}(x)-\theta^{\mu}\zx^{\lambda}\zx^{\nu} R_{~\mu \nu \lambda}^{\rho}(x)\frac{\partial}{\partial \zx^{\rho}}\,,\\
\end{align*}
where $R_{~\mu \nu \lambda}^\rho$ is the Riemann curvature of the Levi-Civita connection (similar expressions can be found in \cite{Hallowell:2008}). The vector fields $\Delta_{(0,1)}$ and $\Delta_{(1,0)}$ have exactly the same local form as on Minkowski space-time.  A direct calculation shows that 
\begin{align*}
[\Delta_{(0,1)}, \nabla_{(1,0)}] = \nabla_{(0,1)}, && [\Delta_{(1,0)}, \nabla_{(0,1)}] = \nabla_{(1,0)} \,.
\end{align*}
where one has to take care with the signs due to the $\Z_2^2$-grading.\par 

The covariant de Rham derivatives are canonical vector fields, once we have fixed a (pseudo-)Riemannian metric. Associated with any (homogeneous) vector field on $\mathcal{M}$ are (local) infinitesimal diffeomorphisms (see Voronov \cite[Section 2.39]{Voronov:1991} for details on standard supermanifolds). For the case at hand, we have a pair of such infinitesimal diffeomorphisms:
\begin{align}\label{eqn:supersym01}
& x^\mu \mapsto x^\mu + \lambda \: \zx^\mu, && \zx^\nu \mapsto \zx^\nu, && \theta^\rho \mapsto \theta^\rho - \lambda  \: \zx^\mu \theta^\nu \Gamma_{\nu \mu}^\rho(x),
\end{align}
and 
\begin{align}\label{eqn:supersym10}
& x^\mu \mapsto x^\mu + \eta \: \theta^\mu, && \zx^\nu \mapsto \zx^\nu - \eta  \: \zx^\mu \theta^\rho \Gamma_{\rho \mu}^\nu(x), && \theta^\rho \mapsto \theta^\rho ,
\end{align}
where $\lambda$ and $\eta$ are ``external'' parameters of $\Z^2_2$-degree $(0,1)$ and $(1,0)$, respectively. Because the parameters carry non-zero degree, such infinitesimal diffeomorphisms can be referred to as \emph{supersymmetries}. However, note that this is different to the standard meaning of a supersymmetry in physics. The action of these supersymmetries on $(p,q)$-forms is, of course, via application of the covariant de Rham differential, i.e., a Lie derivative. Note that these supersymmetries are not  directly associated with (infinitesimal) diffeomorphisms of $M$, but rather come from the larger $\Z_2^n$-manifold structure. We will say that a $(p,q)$-form $\omega^{(p,q)}$ is \emph{$(0,1)$-covariantly constant} if and only if $\nabla_{(0,1)}\omega^{(p,q)}=0$, and similarly a $(p,q)$-form is said to be \emph{$(1,0)$-covariantly constant} if and only if $\nabla_{(1,0)}\omega^{(p,q)}=0$. 
\begin{remark}
For standard differential forms on a manifold, i.e., function on the supermanifold $\Pi \sT M$, we have the infinitesimal diffeomorphism generated by the de Rham differential: 
\begin{align*}
& x^\mu \mapsto x^\mu + \epsilon \: \rmd x^\mu, && \rmd x^\nu \mapsto \rmd x^\nu,
\end{align*}
where $\epsilon$ is a Grassmann odd parameter. As the de Rham differential is a homological vector field, i.e., $[\rmd, \rmd ]= 2 \rmd^2 = 0$, it can be integrated to obtain an odd flow. This produces a canonical action of the Lie supergroup $\R^{0|1}$ on $\Pi \sT M$. Clearly, closed differential forms are the differential forms that are invariant under this action, (see Vaintrob \cite{Vaintrob:1996}). While this should be kept in mind when thinking of bi-differential forms, the covariant de Rham derivatives are not - unless we have a flat manifold - homological vector fields. Thus, we do not expect to have a direct analogue of a  Lie supergroup action for bi-differential forms. 
\end{remark}
\begin{example}
Consider a bi-form  $\omega \in \Omega^{(1,0)}(M)$. Clearly, such a bi-form can be considered as a genuine differential form on $M$. In local coordinates we have that $\omega = \theta^\rho \omega_\rho(x)$. Now, let us consider the pair of supersymmetries: 
 \begin{align*}
        \omega & \mapsto \omega + \lambda \: \nabla_{(0,1)} \omega\\
                & = \theta^\rho \omega_\rho(x) + \lambda \: \theta^\nu \zx^\mu \left (\frac{\partial \omega_\nu(x)}{\partial x^\mu} \:{-} \:\Gamma^\rho_{\mu \nu}\omega_\rho(x) \right)\,,
       \end{align*} 
and
 \begin{align*}
        \omega & \mapsto \omega + \eta \: \nabla_{(1,0)} \omega\\
                & = \theta^\rho \omega_\rho(x) \:{-} \:  \eta \: \theta^\nu \theta^\mu \left (\frac{\partial \omega_\nu(x)}{\partial x^\mu} \: {-} \: \frac{\partial \omega_\mu(x)}{\partial x^\nu}  \right)\,.
       \end{align*} 
In order for $\omega$ to be $(0,1)$-covariantly constant - in the classical framework - it must be parallel (with respect to the Levi-Civita connection). Note that this automatically implied that $\omega$ is closed and so it is also $(1,0)$-covariantly constant. The converse need not be true.  Naturally, the same is true of any $(p,0)$-form and $(0,q)$-form.    
\end{example}
\begin{example}A symmetric rank two covariant tensor can naturally be considered as a $(1,1)$-form. In local coordinates we have that $\omega = \theta^\mu \zx^\nu \omega_{\nu | \mu}(x)$. Now, let us consider the pair of supersymmetries: 
 \begin{align*}
        \omega & \mapsto \omega + \lambda \: \nabla_{(0,1)} \omega\\
                & = \theta^\mu \zx^\nu \omega_{\nu | \mu}(x) + \lambda \: \frac{1}{2!}\theta^\nu \zx^\mu \zx^\rho\left (\nabla_\mu \omega_{\rho|\nu}(x)\:{-}\: \nabla_\rho \omega_{\mu|\nu}(x)\right)\,,
       \end{align*} 
and
 \begin{align*}
        \omega & \mapsto \omega + \eta \: \nabla_{(1,0)} \omega\\
                & = \theta^\mu \zx^\nu \omega_{\nu | \mu}(x) + \lambda \: \frac{1}{2!}\theta^\nu \theta^\mu \zx^\rho\left (\nabla_\nu \omega_{\rho|\mu}(x)\:{-}\: \nabla_\mu \omega_{\rho|\nu}(x)\right)\,.
       \end{align*} 
Then, due to the obvious symmetry, a $(1,1)$-form is invariant under the pair of supersymmetries if and only if
$$\nabla_\mu \omega_{\nu|\rho}(x) \:{-}\:\nabla_\nu \omega_{\mu|\rho}(x) =0\,. $$
As a specific example, the metric tensor $g = \theta^\mu \zx^\nu g_{\nu \mu}$ is invariant under the  supersymmetries as $\nabla_\mu g_{\nu \rho} =0$, i.e., we are using the Levi-Civita connection.  Thus, for Einstein manifolds, e.g., de Sitter and anti de Sitter space-time,  where the Ricci tensor is proportional to the metric, $\textnormal{Ricc} = \theta^\mu \zx^\nu R_{\nu \mu} = k\: \theta^\mu \zx^\nu g_{\nu \mu}$   ($k \in \R^\times$) is invariant under the pair of supersymmetries.  
\end{example}

\begin{example}
The \emph{covariant Riemann tensor} is an example of a $(2,2)$-form on $(M,g)$:
$$R(x, \zx, \theta) = \frac{1}{2! 2!} \theta^ \nu \theta^ \mu \zx^\sigma \zx^\rho \:  R_{\rho \sigma | \mu \nu}(x)\,,$$
here $R_{\rho \sigma | \mu \nu} :=  g_{\rho \lambda}R^\lambda_{~\sigma \mu \nu}$ and  $R_{~\sigma \mu \nu }^\lambda$ is the Riemann curvature of the Levi--Civita connection.  A direct computation shows that the \emph{first Bianchi identity} can be written as
$$\Delta_{(0,1)}R = \frac{1}{3!}\theta^\nu \zx^\rho \zx^\mu \zx^\sigma\big( R_{\nu\sigma| \mu \rho} + R_{\nu \mu | \rho \sigma} + R_{\mu \rho |\sigma \mu}\big) =0\,.$$
Similarly, direct computation shows that the \emph{second Bianchi identity} can be written as 
\begin{align*}
 \nabla_{(0,1)} R  =  \frac{1}{2!3!}\theta^\nu \theta^\rho \zx^\mu \zx^\sigma \zx^\lambda &  \left(  \left( \frac{\partial R_{\sigma \mu |\rho \nu}}{\partial x^\lambda\hfill} ~ {-}~ \Gamma^\omega_{\nu \lambda} R_{\sigma \mu |\rho \omega} ~{-}~ \Gamma^\omega_{\rho \lambda} R_{\sigma \mu |\nu \omega}  \right) \right.
 ~+ ~ \left( \frac{\partial R_{ \mu \lambda |\rho \nu}}{\partial x^\sigma\hfill} ~ {-}~ \Gamma^\omega_{\nu \sigma} R_{\mu \lambda |\rho \omega} ~{-}~ \Gamma^\omega_{\rho \sigma} R_{ \mu \lambda|\nu \omega}  \right)\\
 &~+ ~ \left. \left( \frac{\partial R_{ \lambda \sigma |\rho \nu}}{\partial x^\mu\hfill} ~ {-}~ \Gamma^\omega_{\nu \mu} R_{\lambda \sigma |\rho \omega} ~{-}~ \Gamma^\omega_{\rho \nu} R_{  \lambda \sigma|\nu \omega}  \right) \right) ~=~0\,.
\end{align*}
 Clearly, the second Bianchi identity can also be written as $\nabla_{(1,0)}R =0$ as the covariant Riemann tensor is a $(2,2)$-form. Thus, we see that on \emph{any}   (pseudo-)Riemannian manifold, the  covariant Riemann tensor is a canonical example of a $(0,1)$-covariantly constant and $(1,0)$-covariantly constant $(2,2)$-form. In other words,  the covariant Riemann tensor is preserved under the supersymmetries \eqref{eqn:supersym01} and \eqref{eqn:supersym10}. 
\end{example}

%%%%%%%%%%%%%%%%%%%%%%%%%%%%%%%%%%%%%%%%%%%%%%%%%%%%%%
\section{Vector bundle-valued mixed symmetry tensors}\label{sec:VBvalued}
Consider a vector bundle $\pi : E \rightarrow M$ over a space-time $(M,g)$ of dimension $D$. Similarly, to the previous sections we canonically build a $\Z_2^2$-manifold, but now incorporating the dual vector bundle
$$\mathcal{ME} := \sT M[(0,1)] \times_M \sT M[(1,0)] \times_M E^*[(1,1)]\,.$$
Natural coordinates here are 
$$\big( \underbrace{x^\mu}_{(0,0)}, ~ \underbrace{\zx^\nu}_{(0,1)}, ~ \underbrace{\theta^\rho}_{(1,0)}, ~ \underbrace{z_a}_{(1,1)}  \big)\,,$$ 
where we take the permissible changes of coordinates to be as before, but now including
$$z_{a'} = T_{a'}^{\:\: b}(x) \, z_b\,,$$
which is inherited from the linear changes of fibre coordinates on the vector bundle $E$. Again, we have a split $\Z_2^2$-manifold in this way \cite{Covolo:2016a}. In fact, the reader should note that up to the assignment of the  $\Z_2^n$-grading, we have a decomposed double vector bundle, (see Pradines \cite{Pradines:1974} for the classical description and Voronov \cite{Voronov:2012} for the coordinate description using graded manifolds).  By construction, we have a well-defined $\Z_2^2$-manifold with formal coordinates that are not nilpotent. This means that we must consider formal power series in the coordinates $z$ and not simply polynomials. \par
However, we can - due to the linear nature of the coordinate changes we are allowing - select functions on $\mathcal{ME}$ that are (locally) homogeneous in $z$. We then define \emph{$E$-valued bi-forms} in the following way:
$$\Omega^{(p,q)}(M,E):=  C^\infty(\mathcal{ME})_{(p,q,1)}\,.$$
In terms of local coordinates,  any $\omega^{(p,q)} \in \Omega^{(p,q)}(M,E)$ has the local form
$$\omega^{(p,q)}(x, \zx, \theta,z) = \frac{1}{p! q!} \: \theta^{\nu_1} \cdots \theta^{\nu_p} \zx^{\mu_1} \cdots \zx^{\mu_q}  \: \omega^a_{\mu_q \cdots \mu_1 | \nu_q \cdots \nu_1}(x)z_a\,.$$ 
Naturally, we have the identification $\Omega^{(0,0)}(M,E) \simeq \Sec(E)$. By considering all the possible degrees we obtain the vector space of all $E$-valued bi-forms:
$$\Omega(M,E) :=  \bigoplus_{(p,q) \in \N\times \N}^{(D,D)} \Omega^{(p,q)}(M,E),$$
which has the obvious (left) module structure over the algebra of bi-forms $\Omega(M)$. 

If we specify a  linear connection on $E$, then we can construct a pair of  \emph{(fully) covariant de Rham derivatives}
\begin{align*}
& \nabla_{(0,1)} := \zx^\mu \nabla_\mu = \zx^\mu \frac{\partial}{\partial x^\mu } {-} \zx^\mu\theta^\nu \Gamma_{\nu \mu}^{\rho} \frac{\partial}{\partial \theta^\rho} + \zx^\nu (\mathbb{A}_\mu)_a^{\:\: b}z_b \frac{\partial}{\partial z_a}\,,\\
& \nabla_{(1,0)} := \theta^\mu \nabla_\mu = \theta^\mu \frac{\partial}{\partial x^\mu } {-} \zx^\mu\theta^\nu \Gamma_{\nu \mu}^{\rho} \frac{\partial}{\partial \zx^\rho} + \theta^\mu  (\mathbb{A}_\mu)_a^{\:\: b}z_b \frac{\partial}{\partial z_a}\,,
\end{align*}
where $(\mathbb{A}_\mu)_a^{\:\: b}$ are the components of the (local) connection one-form associated with the linear connection.
\begin{example}
If we take $E = \sT M$, then we obtain \emph{tangent bundle-valued bi-forms}. The degree $(1,1)$ coordinates we can view as ``momenta'' as they correspond, up to the grading, with fibre coordinates on $\sT^* M$. naturally, the  Levi-Civita connection gives rise to a pair of  covariant de Rham derivatives.
\end{example}
\begin{example}
If we consider  the real spinor bundle over a (pseudo-)Riemannian spin manifold $(M,g)$, which we denote as $\Sigma M$, then we naturally have the notion of \emph{spinor-valued bi-forms}.  Moreover, the Levi-Civita connection induces a linear connection on $\Sigma M$ (see \cite[Chapter II, \S 4]{Lawson:1989}) and so canonically we have a pair of (fully) covariant de Rham derivatives acting on spinor-valued bi-forms. 
\end{example}
\begin{example}
Consider a line bundle $\pi : L \rightarrow M$, then we can build $\mathcal{ML}$ as described above. The transformation law for the degree $(1,1)$ coordinate is of the form $z' = \phi^{-1}(x) \; z$, where $\phi(x)$ are the transition functions on $L$. As a specific example, we can consider the density bundle and so we have the natural notion of \emph{twisted bi-forms}. Moreover, as the density bundle is trivial, we can use the covariant de Rham derivatives as defined in the previous section, i.e., we can use the trivial connection on $L$. 
\end{example}
Again, we can consider a pair of ``supersymmetries'' along the same lines as \eqref{eqn:supersym01} and \eqref{eqn:supersym10}, but now with the additional terms
\begin{align*}
& z_a \mapsto z_a + \lambda \: \zx^\mu (\mathbb{A}_\mu)_a^{\:\: b}(x) z_b\,,&
&  \textnormal{and} & z_a \mapsto z_a + \eta \: \theta^\mu (\mathbb{A}_\mu)_a^{\:\: b}(x) z_b\,.
\end{align*}

\section{Concluding Remarks}\label{sec:Concl}
As remarked in the introduction, differential forms on a manifold $M$ are naturally understood as functions of the antitangent bundle $\Pi \sT M$, which itself canonically  comes equipped with the de Rham differential, here understood as a homological vector field.  Similarly, bi-forms on a (pseudo-)Riemannian manifold $(M,g)$, are naturally understood as functions on the $\Z_2^2$-manifold $\sT M[(0,1)]\times_M \sT M[(1,0)]$, which canonically comes equipped with the odd vector fields (generally, non-homological)  $\nabla_{(0,1)}$ and $\nabla_{(1,0)}$. We have shown that there is a natural pair of ``supersymmetries'' associated with these covariant de Rham derivatives. Moreover, the metric tensor and the covariant Riemann tensor are invariant under these transformations. This, in turn, implies that for Einstein manifolds, the Ricci tensor is also invariant under these supersymmetries. Similar statements can be made for more general multi-forms. We have also shown that this geometric framework can be extended to cover bi-forms with values in vector bundles,  which include spinor-valued and twisted bi-forms. While the goals of this paper have been modest, we hope that the observations made here will prove useful for future studies of mixed symmetry tensors.  In particular, the geometric aspects of mixed symmetry tensors seem to have been largely missed within the mathematics literature.

%%%%%%%%%%%%%%%%%%%%%%%%%%%%%%%%%%%%%%%%%%%%%%%%%%%%%%
\section*{Acknowledgements}
We thank Norbert Poncin for many discussions relating to $\Z_2^n$-geometry. We also thank Richard Szabo for his interest in earlier versions of this note.  A special thank you goes to Andrea Campoleoni and Andrew Waldron for their help with navigating some of the physics literature on mixed symmetry tensors.

%%%%%%%%%%%%%%%%%%%%%%%%%%%%%%%%%%%%%%%%%%%%%%%%%%%%%%

\end{document}